\documentclass[aps,pra,showpacs,twocolumn]{revtex4-1}
\usepackage{amssymb}
\usepackage{amsmath}
\usepackage{graphicx}
\usepackage{epsfig}
\begin{document}

\title{Conventional quantum phase transition driven by complex parameter in
non-Hermitian $\mathcal{PT}$-symmetric Ising model}
\author{C. Li, G. Zhang, X. Z. Zhang and Z. Song}
\email{songtc@nankai.edu.cn}
\affiliation{School of Physics, Nankai University, Tianjin 300071, China}

\begin{abstract}
A conventional quantum phase transition (QPT) can be accessed by varying a
real parameter at absolute zero temperature. Motivated by the discovery of
the pseudo-Hermiticity of non-Hermitian systems, we explore the QPT in
non-Hermitian $\mathcal{PT}$-symmetric Ising model, which is driven by a
staggered complex transverse field. Exact solution shows that the Laplacian
of the groundstate energy density, with respect to real and imaginary
components of the transverse field, diverges on the boundary in the complex
plane. The phase diagram indicate that the imaginary transverse field has
the effect of shrinking the paramagnet phase. In addition, we also
investigate the connection between the geometric phase and the QPT.
\end{abstract}

\maketitle


\section{Introduction}

\label{sec_intro}

Quantum phase transitions (QPTs) happen at zero temperature\ when
physical parameters are changed, inducing\ dramatic changes in the
ground-state properties \cite{S.Sachdev}. So far, these system-specific
parameters are required to be real, which can be a magnetic field in spin
systems \cite{R.Coldea,Sadler}, the intensity of a laser beam in cold-atom
simulators of Hubbard-like models \cite{M.Greiner}, the dopant concentration
in high-Tc superconductors \cite{P.A.Lee}, etc.

With the discovery that a non-Hermitian Hamiltonian having simultaneous
parity-time ($\mathcal{PT}$) symmetry has a real spectrum \cite{Bender 98},
there has been an intense effort to establish a $\mathcal{PT}$-symmetric
quantum theory as a complex extension of the conventional quantum mechanics
\cite{Bender 99,Dorey 01,Bender 02,A.M43,A.M,A.M36,Jones,AM}. Motivated by
the pseudo-Hermiticity of non-hermitian systems, it is natural to ask
whether a complex parameter can drive a QPT. Here the QPT does not include
the phase transition in the context of the complex quantum mechanics, which
happens when the reality of the spectrum does not ensure diagonalizability,
associating with spontaneous $\mathcal{PT}$-symmetry breaking. As system
parameter varying, a sudden changes in the eigenstate rather than specifying
the ground state and the critical point is referred as exceptional point.

In traditional condensed matter approaches, for the case of second-order
QPTs, the critical point is identified by the divergence of the second-order
derivative of the ground state density, with respect to the real parameter.
It is interesting to investigate the QPT of a non-Hermitian system, where
the transition is driven by the competition between real and imaginary parts
of parameter.

In this paper, we explore the QPT in non-Hermitian $\mathcal{PT}$-symmetric
Ising model, which is driven by a staggered complex transverse field. Exact
solution shows that the Laplacian of the groundstate energy density, with
respect to real and imaginary components of the transverse field, diverges
on the boundary in the complex plane. The phase diagram indicates that the
imaginary transverse field has the effect of shrinking the paramagnet phase.
We also investigate the connection between the geometric phase and the QPT\
in the present model as that in the study of the conventional quantum spin
model. We find that the phase boundary can be identified by divergence of
Berry curvature density.

This paper is organized as follows. In Section \ref{sec_model}, we present
the model Hamiltonian, where the Ising ring is subjected to a staggered
complex magnetic field. The exact solution allows us to identify the role of
the complex field. In Section \ref{sec_Phase diagram}, we investigate the
phase diagram by the Laplacian of the groundstate energy density. Section %
\ref{sec_Berry curvature} is devoted to another characterization of the QPT
in terms of the geometric phase of the ground state. Finally, we give a
summary and discussion in Section \ref{sec_summary}.

\section{Model and solution}

\label{sec_model}We consider a non-Hermitian one-dimensional spin-$1/2$
Ising model in a complex staggered\ transverse magnetic field on a $2N$-site
lattice. The system is modeled by the following Hamiltonian%
\begin{equation}
\mathcal{H}\text{ }=-J\underset{j=1}{\overset{2N}{\sum }}\left( \sigma
_{j}^{z}\sigma _{j+1}^{z}+g_{_{j}}\sigma _{j}^{x}\right) ,  \label{H_general}
\end{equation}%
where $g_{_{j}}=\eta +\left( -1\right) ^{j+1/2}\xi $\ with $\eta $ and $\xi $%
\ being real numbers. Here $\sigma _{j}^{\lambda }$ ($\lambda =x,$ $z$) are
the Pauli operators on site $j$, and satisfy the periodic boundary condition
$\sigma _{j}^{\lambda }\equiv \sigma _{j+2N}^{\lambda }$. We note that the
non-Hermiticity of the Hamiltonian arises from complex staggered\ transverse
magnetic field. One can define a parity operator $\mathcal{P}$\ which has
the function
\begin{equation}
\mathcal{P}\sigma _{j}^{\lambda }\mathcal{P}^{-1}\equiv \sigma
_{2N+1-j}^{\lambda },  \label{P}
\end{equation}%
and a time reversal operator $\mathcal{T}$ which has the function%
\begin{equation}
\mathcal{T}\sigma _{j}^{\lambda }\mathcal{T}^{-1}\equiv \left\{
\begin{array}{c}
-\sigma _{j}^{\lambda },\lambda =y \\
\sigma _{j}^{\lambda },\lambda =x,z%
\end{array}%
\right. ,  \label{T}
\end{equation}%
It turns out that, for nonzero $\xi $, we have $\left[ \mathcal{P},\mathcal{H%
}\right] \neq 0$ and $\left[ \mathcal{T},\mathcal{H}\right] \neq 0$, but

\begin{equation}
\left[ \mathcal{PT},\mathcal{H}\right] =0,  \label{PT}
\end{equation}%
where the antilinear time reversal operator $\mathcal{T}$ has the function $%
\mathcal{T}i\mathcal{T=-}i$, i.e., the Hamiltonian $\mathcal{H}$ is
parity-time ($\mathcal{PT}$) reversal invariant.

\begin{figure}
\includegraphics[ bb=59 366 550 793, width=0.35\textwidth, clip]{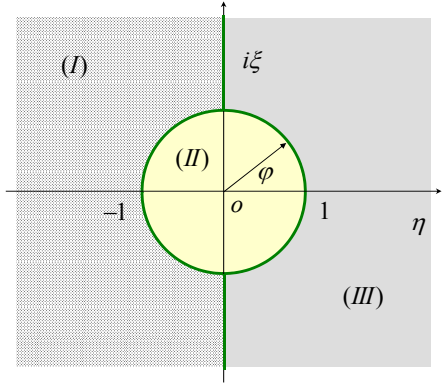}
\caption{(Color online) Phase diagram for the ground state of the Ising ring
in a a staggered complex transverse field. The heavy lines represent the
boundary which separates three quantum phases. Phases \textrm{I} and \textrm{III} are
paramagnet, while \textrm{II} is ferromagnet.} \label{fig1}
\end{figure}

Now we consider the solution of the non-Hermitian Hamiltonian of Eq. (\ref%
{H_general}). We start by taking the Jordan-Wigner transformation \cite%
{P.Jordan}%
\begin{eqnarray}
\sigma _{j}^{+} &=&\prod\limits_{l<j}\left( 1-2c_{l}^{\dag }c_{l}\right)
c_{j}, \\
\sigma _{j}^{-} &=&\prod\limits_{l<j}\left( 1-2c_{l}^{\dag }c_{l}\right)
c_{j}^{\dag }, \\
\sigma _{j}^{x} &=&1-2c_{j}^{\dag }c_{j}, \\
\sigma _{j}^{z} &=&-\prod\limits_{l<j}\left( 1-2c_{l}^{\dagger }c_{l}\right)
\left( c_{j}+c_{j}^{\dagger }\right) ,
\end{eqnarray}%
to replace the Pauli operators by the fermionic operators $c_{j}$. Likewise,
the parity of the number of fermions%
\begin{equation}
\Pi =\prod_{l=1}^{2N}\left( \sigma _{l}^{x}\right) =\left( -1\right) ^{N_{p}}
\end{equation}%
\bigskip is a conservative quantity, i.e., $\left[ \mathcal{H},\Pi \right]
=0 $, where $N_{p}=\sum_{j=1}^{2N}c_{j}^{\dag }c_{j}$. Then the Hamiltonian (%
\ref{H_general}) can be rewritten as%
\begin{equation}
\mathcal{H}=\sum_{\zeta =+,-}P_{\zeta }\mathcal{H}_{\zeta }P_{\zeta },
\end{equation}%
where
\begin{equation}
P_{\zeta }=\frac{1}{2}\left( 1+\zeta \Pi \right)
\end{equation}%
is the projector on the subspaces with even ($\zeta =+$) and odd ($\zeta =-$%
) $N_{p}$. The Hamiltonian in each invariant subspaces has the form%
\begin{eqnarray}
\mathcal{H}_{\zeta } &=&-J\sum\limits_{j=1}^{2N-1}\left( c_{j}^{\dag
}c_{j+1}+c_{j+1}^{\dag }c_{j}+c_{j}^{\dag }c_{j+1}^{\dag
}+c_{j+1}c_{j}\right)  \notag \\
&&+J\zeta \left( c_{2N}^{\dag }c_{1}+c_{1}^{\dag }c_{2N}+c_{2N}^{\dag
}c_{1}^{\dag }+c_{1}c_{2N}\right)  \notag \\
&&-Jg_{j}\sum\limits_{j=1}^{2N-1}\left( 1-2c_{j}^{\dagger }c_{j}\right)
\label{H_sub}
\end{eqnarray}%
taking the Fourier transformation%
\begin{equation}
c_{j}=\frac{1}{\sqrt{N}}\sum\limits_{k_{\zeta }}e^{ik_{\zeta }j}\left\{
\begin{array}{cc}
\alpha _{k_{\zeta }}, & \text{even }j \\
\beta _{k_{\zeta }}, & \text{odd }j%
\end{array}%
\right. ,
\end{equation}%
for the Hamiltonians $\mathcal{H}_{\zeta }$, we have%
\begin{eqnarray}
\mathcal{H}_{\zeta } &=&-J\sum_{k_{\zeta }}H_{k_{\zeta }} \\
H_{k_{\zeta }} &=&\left( e^{ik}+1\right) \alpha _{k_{\zeta }}^{\dagger
}\beta _{k_{\zeta }}+\left( e^{ik}-1\right) \alpha _{k_{\zeta }}^{\dagger
}\beta _{-k_{\zeta }}^{\dagger }+\mathrm{H.c.}  \notag \\
&&+2\eta -2\left( \eta +i\xi \right) \alpha _{k_{\zeta }}^{\dagger }\alpha
_{k_{\zeta }} \\
&&-2\left( \eta -i\xi \right) \beta _{k_{\zeta }}^{\dag }\beta _{k_{\zeta }}
\end{eqnarray}%
where the momentum $k_{\zeta }$ are defined as $k_{+}=2\left( m+1/2\right)
\pi /N$, $k_{-}=2m\pi /N$, $m=0,1,2,...,N-1$, respectively.

In the following, we focus on the subspace with $\zeta =+$ since it turns
out that the ground state lies in this sector in the thermodynamic limit. We
will neglect the subscript $\zeta $\ in $\mathcal{H}_{\zeta }$\ and $%
k_{\zeta }$. In order to diagonalize the Hamiltonian $\mathcal{H}$, we
introduce the composite operators $\overline{\Lambda }_{n}^{k}$ ($n\in \left[
1,6\right] $), defined as
\begin{eqnarray}
\overline{\Lambda }_{n}^{k} &=&\frac{1}{\Omega _{n}^{k}}[e^{ik/2}\alpha
_{k}^{\dagger }\beta _{-k}^{\dagger }-e^{-ik/2}\beta _{k}^{\dagger }\alpha
_{-k}^{\dagger }  \notag \\
&&+2\cos \left( k/2\right) \left( \frac{\alpha _{k}^{\dagger }\alpha
_{-k}^{\dagger }}{\epsilon _{n}^{k}+i2\xi }+\frac{\beta _{k}^{\dagger }\beta
_{-k}^{\dagger }}{\epsilon _{n}^{k}-i2\xi }\right)  \notag \\
&&-2i\sin \left( k/2\right) \left( \frac{1}{\epsilon _{n}^{k}+2\eta }\alpha
_{k}^{\dagger }\beta _{k}^{\dagger }\alpha _{-k}^{\dagger }\beta
_{-k}^{\dagger }+\frac{1}{\epsilon _{n}^{k}-2\eta }\right) ],  \notag \\
&&\left( n\in \left[ 1,5\right] \right) ,
\end{eqnarray}%
where the normalization factor is%
\begin{eqnarray}
\left( \Omega _{n}^{k}\right) ^{2} &=&2+\frac{4\cos ^{2}\left( k/2\right) }{%
\left( \epsilon _{n}^{k}+2i\xi \right) ^{2}}+\frac{4\cos ^{2}\left(
k/2\right) }{\left( \epsilon _{n}^{k}-2i\xi \right) ^{2}}  \notag \\
&&+\frac{4\sin ^{2}\left( k/2\right) }{\left( \epsilon _{n}^{k}+2\eta
\right) ^{2}}+\frac{4\sin ^{2}\left( k/2\right) }{\left( \epsilon
_{n}^{k}-2\eta \right) ^{2}},
\end{eqnarray}%
and%
\begin{equation}
\overline{\Lambda }_{6}^{k}=\frac{1}{\sqrt{2}}\left( e^{ik/2}\alpha
_{k}^{\dagger }\beta _{-k}^{\dagger }+e^{-ik/2}\beta _{k}^{\dagger }\alpha
_{-k}^{\dagger }\right) .
\end{equation}%
Here coefficients $\epsilon _{n}^{k}$\ are defined as%
\begin{eqnarray}
\epsilon _{1}^{k} &=&\sqrt{2r^{2}\cos \left( 2\varphi \right) +2\sqrt{%
r^{4}-2r^{2}\cos k+1}+2},  \notag \\
\epsilon _{3}^{k} &=&\sqrt{2r^{2}\cos \left( 2\varphi \right) -2\sqrt{%
r^{4}-2r^{2}\cos k+1}+2},  \label{spec} \\
\epsilon _{2}^{k} &=&-\epsilon _{1}^{k}\text{, }\epsilon _{4}^{k}=-\epsilon
_{3}^{k}\text{, }\epsilon _{5}^{k}=\epsilon _{6}^{k}=0\text{,}  \notag
\end{eqnarray}%
where we parameterize the complex field in terms of the polar radius and
angle

\begin{equation}
r=\sqrt{\eta ^{2}+\xi ^{2}}\text{ and }\tan \varphi =\xi /\eta ,
\end{equation}%
as shown in figure \ref{fig1}. Similarly, we also introduce the composite
operators $\Lambda _{n}^{k}$\ by the following procedure%
\begin{equation}
\Lambda _{n}^{k}=\left[ \overline{\Lambda }_{n}^{k}\left( \xi \rightarrow
-\xi \right) \right] ^{\dag },
\end{equation}%
which will be used to construct the biorthogonal set together with $%
\overline{\Lambda }_{n}^{k}$. Straightforward calculation shows that%
\begin{equation}
\left\langle 0\right\vert \Lambda _{m}^{k^{\prime }}\overline{\Lambda }%
_{n}^{k}\left\vert 0\right\rangle =\delta _{mn}\delta _{kk^{\prime }},
\label{bio}
\end{equation}%
and%
\begin{eqnarray}
h_{k}\overline{\Lambda }_{n}^{k}\left\vert 0\right\rangle &=&2\epsilon
_{n}^{k}\overline{\Lambda }_{n}^{k}\left\vert 0\right\rangle , \\
\left\langle 0\right\vert \Lambda _{m}^{k}h_{k}^{\dag } &=&\left\langle
0\right\vert \Lambda _{m}^{k}2\epsilon _{n}^{k},
\end{eqnarray}%
where $h_{k}=H_{k}+H_{-k}$ and $\left\vert 0\right\rangle $\ is the vaccum
of fermion operator $c_{j}$,\ i.e., $c_{j}\left\vert 0\right\rangle =0$. Eq.
(\ref{bio}) indicates the biorthogonality relation between the eigenstates
of $H_{k}$. Accordingly, all the eigenstates of $\mathcal{H}$\ can be
constructed by the product of complete biorthogonal basis set $\left\{
\overline{\Lambda }_{n}^{k}\left\vert 0\right\rangle \right\} $ as the form $%
\prod_{\left\{ k,n\right\} }\overline{\Lambda }_{n}^{k}\left\vert
0\right\rangle $. It can be seen that part of eigenvalues of $\mathcal{H}$\
can be complex, which does not affect our investigation.

\begin{figure}
\includegraphics[ bb=6 44 536 540, width=0.33\textwidth, clip]{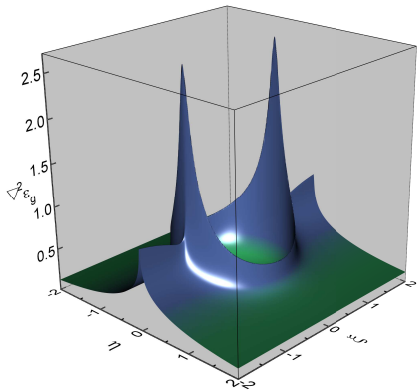}
\caption{(Color online) The Laplacian of the groundstate energy density $\protect\varepsilon _{g}$ as a function of the field for the case $N=300$. The peaks mark the clear regions of criticality.} \label{fig2}
\end{figure}

In the following analysis, we will focus on the ground state (or the eigen
state with the lowest real eigenvalue) of the Hamiltonian. The
ground state of $\mathcal{H}$ can be constructed as the form
\begin{equation}
\left\vert G\right\rangle =\prod_{0<k<\pi }\overline{\Lambda }%
_{1}^{k}\left\vert 0\right\rangle ,  \label{GS}
\end{equation}%
with the eigenvalue%
\begin{equation}
E_{g}=-2\sum\limits_{k}\epsilon _{1}^{k}.  \label{E_g}
\end{equation}%
where $k=2\pi \left( m+1/2\right) /N$, $m=0,1,2,\ldots ,N/2-1$. Accordingly
the bra ground state can be expressed as the form%
\begin{equation}
\left\langle \overline{G}\right\vert =\left\langle 0\right\vert
\prod_{0<k<\pi }\Lambda _{1}^{k}.
\end{equation}

It is worth stressing that the diagonalization procedure we have used here
is a little different from the Bogoliubov transformation which is applied
for the standard transverse-field Ising model \cite{Lieb}. Here $\Lambda
_{n}^{k}$ and $\overline{\Lambda }_{n}^{k}$ are composite operators, which
do not obey the canonical commutation relations as the fermion operators in
the Bogoliubov transformation, but the biorthonormal relation in Eq. (\ref%
{bio}).
\begin{figure*}
\includegraphics[ bb=29 52 500 538, width=0.32\textwidth, clip]{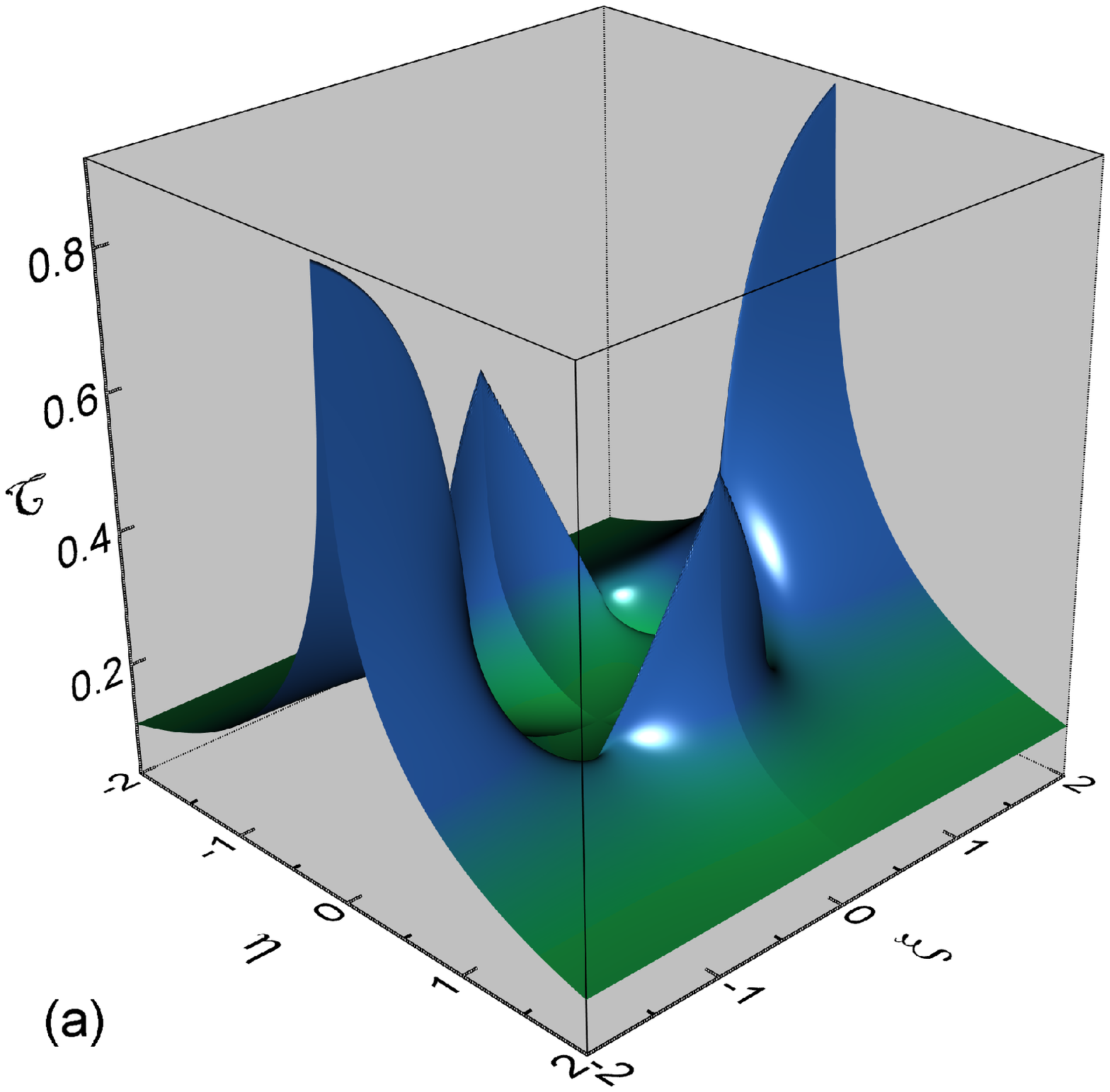} %
\includegraphics[ bb=29 52 505 542, width=0.32\textwidth, clip]{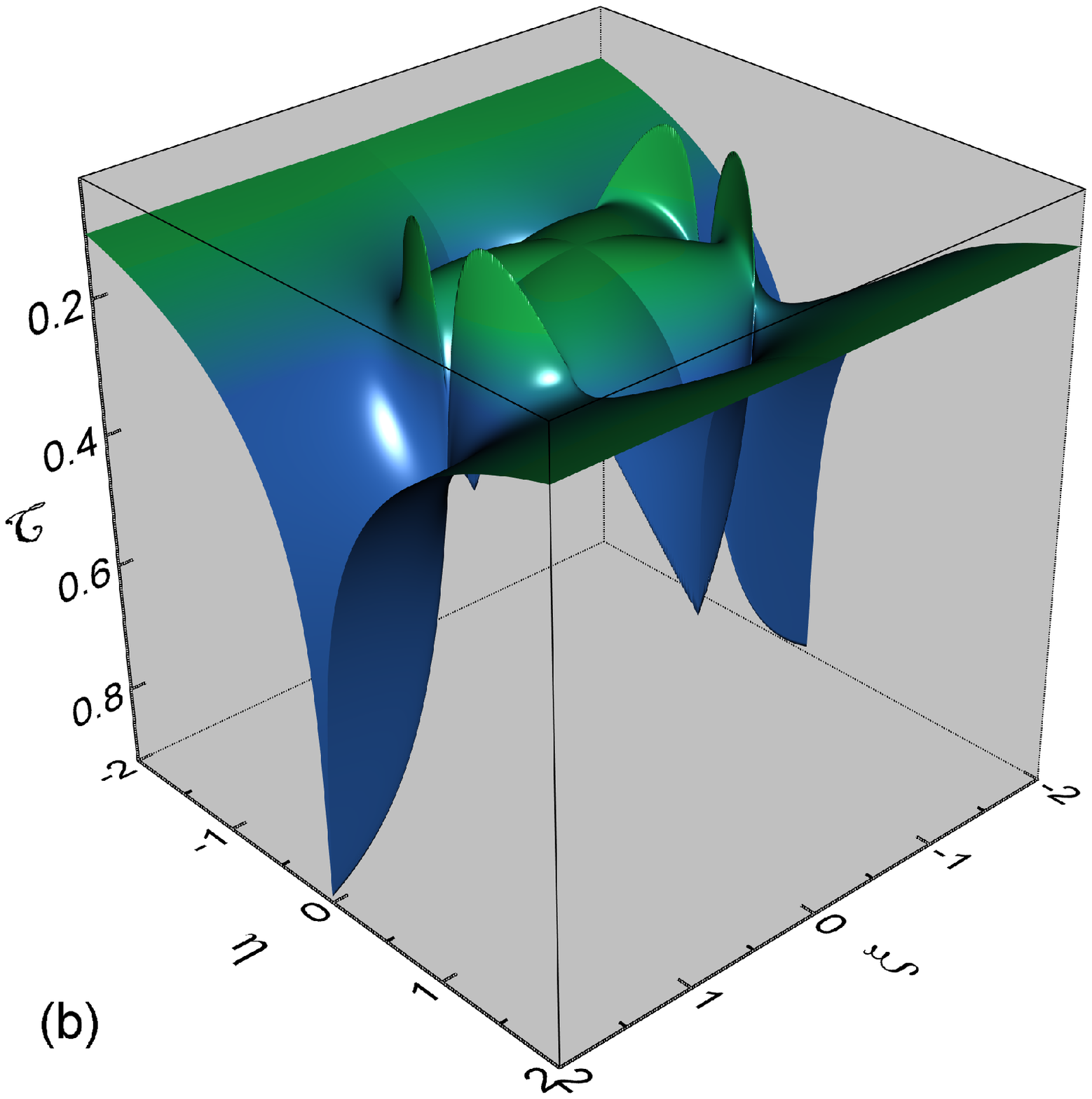} %
\includegraphics[ bb=0 0 519 538, width=0.32\textwidth, clip]{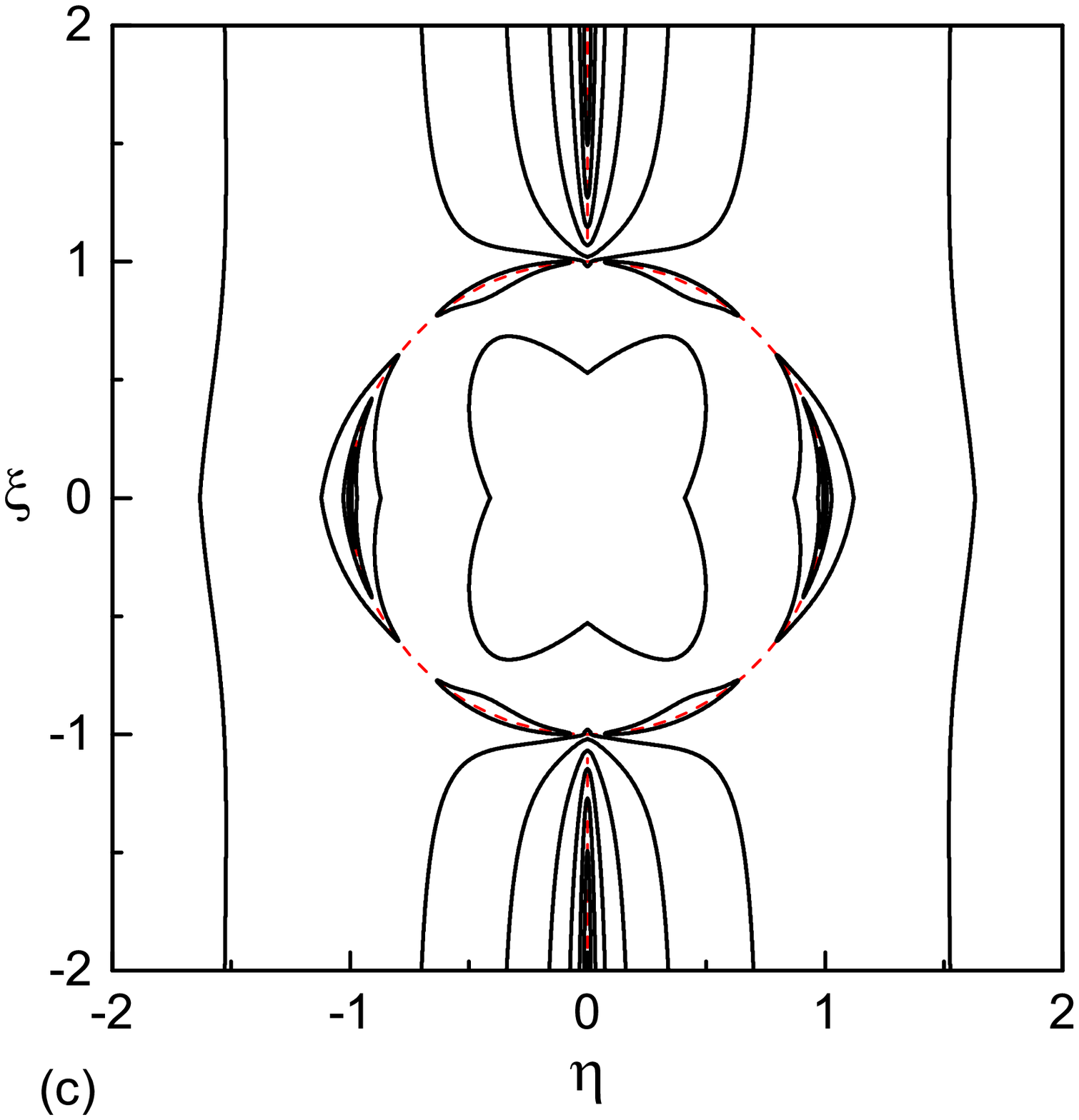}
\caption{(Color online) (a) The curvature density $\mathcal{C}$ as a
function of the field for the Hamiltonians with the parameters of (a) Eq. (%
\protect\ref{function1}), and $N=300$. The plots (b) is the inversion of
(a). The dips and peaks indicate the quasi-critical lines. (c) Contour map
of $\mathcal{C}$. The red dashed lines indicate the phase diagram in Fig.
\protect\ref{fig1}.}
\label{fig3}
\end{figure*}

\section{Phase diagram}

\label{sec_Phase diagram}In this section, we will investigate the phase
diagram of the Hamiltonian (\ref{H_general}) based on the solutions. In all
previous study for a non-Hermitian system, the term phase diagram has a
little different meaning from that of a Hermitian system. It usually
represents the region in which the non-Hermitian Hamiltonian has full real
spectrum or not (as examples of non-Hermitian quantum spin systems, see Ref.
\cite{zxzspin,zxzspin2}), rather than the quantum phase transition in a
Hermitian system\cite{S.Sachdev}, which specifies the sudden change of the
ground state as a real parameter varies. However, in this paper, we are
interested in the sudden change of the state $\left\vert G\right\rangle $ as
the complex field $g_{_{j}}$ varies. The aim of this work is to investigate
the conventional QPT occurs in the present non-Hermitian spin system.

To this aim, we investigate the value of $\epsilon _{1}^{k}$ at $k=0$, which
is
\begin{equation}
\epsilon _{1}^{0}=\sqrt{2r^{2}\cos \left( 2\varphi \right) +2\left\vert
r^{2}-1\right\vert +2}.
\end{equation}%
We note that $\epsilon _{1}^{0}$ has a discontinuous derivative at $r=1$. On
the other hand, for $r>1$, we have%
\begin{equation}
\epsilon _{1}^{0}=2r\left\vert \cos \varphi \right\vert ,
\end{equation}%
which indicates a discontinuous derivative at $\eta =0$. These boundary
lines separate the ground state into three phases as illustrated in Fig. \ref%
{fig1}, with I and III being paramagnet, II being ferromagnet. Quantum phase
transition takes place at the critical value $r=1$ and $\eta =0$\ ($r>1$)\
of external field. When $\eta \gg 1$ ($-\eta \gg 1$), then the ground state
is a paramagnet $\prod_{l=1}^{2N}\left\vert \rightarrow \right\rangle _{l}$ (%
$\prod_{l=1}^{2N}\left\vert \leftarrow \right\rangle _{l}$) with all spins
polarized up (down) along the $x$ axis. In this limit case, the imaginary
field $\xi $\ has no contribution to the groundstate energy. On the other
hand, when $r=0$, then there are two degenerate ferromagnetic ground states
with all spins pointing either up or down along the $z$ axis: $%
\prod_{l=1}^{2N}\left\vert \uparrow \right\rangle _{l}$ or $%
\prod_{l=1}^{2N}\left\vert \downarrow \right\rangle _{l}$. It can be seen
that nonzero imaginary field $\xi $\ seems to suppress the influence of the
Ising term $\sum \sigma _{j}^{z}\sigma _{j+1}^{z}$, shrinking the
ferromagnetic phase area in the $\eta $ axis.

Now we further investigate the behavior of groundstate energy density $%
\varepsilon _{g}=E_{g}/\left( 2N\right) $ as function of $\eta $ and $\xi $
in the following two cases: i) for $\eta \neq 0$ when $r$\ crosses $1$, ii) $%
\left\vert \xi \right\vert \gg 1$, when $\eta $ crosses $0$. To characterize
this situation, we calculate the Laplacian of $\varepsilon _{g}$
\begin{equation}
\triangledown ^{2}\varepsilon _{g}=\frac{\partial ^{2}\varepsilon _{g}}{%
\partial \xi ^{2}}+\frac{\partial ^{2}\varepsilon _{g}}{\partial \eta ^{2}},
\end{equation}%
which will reduce to second derivative of the groundstate energy density of
the standard transverse-field Ising model \cite{S.Sachdev}\ with respect to
the transverse field $\eta $\ when we take $\xi =0$.\ The physical meaning
of $\triangledown ^{2}\varepsilon _{g}$\ will be given in the next section.

i) The case of $\eta \neq 0$ and $r\rightarrow 1$. The main contribution of\
the Laplacian of $\varepsilon _{g}$ near this boundary can be expressed as%
\begin{equation}
\triangledown ^{2}\varepsilon _{g}\sim \underset{k}{\sum }\frac{2\sqrt{2}%
r\sin ^{2}k}{N\epsilon _{1}^{k}\left( r^{4}-2r^{2}\cos k+1\right) ^{3/2}}
\end{equation}%
In the thermodynamic limit we have%
\begin{equation}
\triangledown ^{2}\varepsilon _{g}\approx \int_{0}^{\pi }\digamma \left(
k\right) \text{d}k,
\end{equation}%
where the integrand is defined as%
\begin{equation}
\digamma \left( k\right) =\frac{\sqrt{2}r\sin ^{2}k}{\pi \epsilon
_{1}^{k}\left( r^{4}-2r^{2}\cos k+1\right) ^{3/2}}.
\end{equation}

We are interested in the divergent behavior when $r^{2}\sim 1$.\ We note
that the main contribution comes from $k\in \left[ 0,\delta \right] $ with $%
\delta \ll \pi $.\ Then we have%
\begin{equation}
\triangledown ^{2}\varepsilon _{g}\approx \int_{0}^{\delta }\digamma \left(
k\right) \text{d}k\approx -\frac{\sqrt{2}}{\pi \left\vert \cos \varphi
\right\vert }\ln \left\vert r-1\right\vert .
\end{equation}

ii) The case of $\left\vert \xi \right\vert \gg 1$ and $\eta \rightarrow 0$.
By the similar analysis as above, we have%
\begin{equation}
\triangledown ^{2}\varepsilon _{g}\approx \frac{\sqrt{2}}{\pi }%
\int_{0}^{\delta }\frac{1}{\epsilon _{1}^{k}}dk\approx -\frac{\sqrt{2}}{\pi }%
\ln \left\vert \eta \right\vert .
\end{equation}%
Then we conclude that the Laplacian of $\varepsilon _{g}$ is divergent at
the boundary illustrated in Fig. \ref{fig1}.

It is crucial to stress that such phase separation does not arise from the
breaking of the symmetries defined by Eqs. (\ref{PT}) as that in
non-Hermitian systems have been investigated heretofore. It is easy to check
that

\begin{equation}
\mathcal{PT}\left\vert G\right\rangle =\pm \left\vert G\right\rangle
\end{equation}%
which indicates that the ground state have $\mathcal{PT}$ symmetries in all
region, due to the relations

\begin{eqnarray}
\mathcal{PT}\alpha _{k}^{\dagger }\left( \beta _{k}^{\dagger }\right) \left(
\mathcal{PT}\right) ^{-1} &=&-e^{-2ik}\beta _{k}^{\dagger }\alpha
_{k}^{\dagger } \\
\mathcal{PT}\left( \overline{\Lambda }_{1}^{k}\right) \left( \mathcal{PT}%
\right) ^{-1} &=&-\overline{\Lambda }_{1}^{k}
\end{eqnarray}%
We conclude this section by presenting the numerical simulation of $%
\triangledown ^{2}\varepsilon _{g}$\ as function of the complex field for
finite $N$ system. In Fig. \ref{fig2} we plot the Laplacian of $\varepsilon
_{g}$ for the case $N=300$. We observe that the regions of criticality are
clearly marked by a sudden increase of the value of $\triangledown
^{2}\varepsilon _{g}$. As before in the Hermitian system, we ascribe this
type of behavior to a dramatic change in the structure of the ground state
of the system while undergoing QPT.

\section{Berry curvature}

\label{sec_Berry curvature}

\begin{figure}[tbp]
\includegraphics[ bb=20 64 523 536, width=0.33\textwidth, clip]{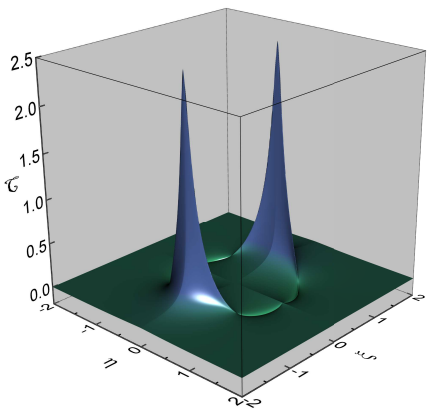}
\caption{(Color online) The curvature density $\mathcal{C}$ as a function of
the field for the Hamiltonians with the parameters of Eq. (\protect\ref{function2}), and $N=300$. The peaks only indicate the quasi-critical lines
of the circle in the Fig. \protect\ref{fig1}, but not the straight lines on
the $\protect\xi $\ axis.} \label{fig4}
\end{figure}

In the present section, we study the geometric phase for the ground state in
the vicinity of the quantum phase boundary. In the realm of traditional
quantum mechanics, geometric phase has been introduced to analyze the
quantum phase transitions of the XY model \cite{Carollo,Zhu1,Zhu2}, and much
effort has been devoted to various Hermitian many-body systems \cite{G.
Chen,X. X. Yi,Zanardi,Chenshu,Tao Liu,Ka-Di Zhu,WXG,Fulibin,Ling-Bao Kong}.
A natural question is whether or not the geometric phase of the ground state
in the present model can be utilized to characterize the quantum phase
boundary. With particular form of parameter dependence on the external
field, we will show that the boundary corresponds to the divergence of the
Berry curvature.

We consider a family of the Hamiltonians that can be obtained by applying a
rotation of $\theta _{a}$ and $\theta _{b}$\ around the $x$-direction for
spins in sublattice A and B, respectively. We have

\begin{equation}
H(\theta _{a},\theta _{b})=R(\theta _{a},\theta _{b})HR^{\dag }(\theta
_{a},\theta _{b})
\end{equation}%
with the unitary operator%
\begin{equation}
R(\theta _{a},\theta _{b})=\prod_{l_{a},l_{b}}e^{i\sigma _{l_{a}}^{x}\theta
_{a}}e^{i\sigma _{l_{b}}^{x}\theta _{b}}.
\end{equation}%
The family of Hamiltonians that is parameterized by real $\theta _{a},\theta
_{b}$ is clearly isospectral and, therefore, the critical behavior is
independent from $\theta _{a},\theta _{b}$. In addition, due to its bilinear
form, $H(\theta _{a},\theta _{b})$ is $\pi $-periodic in $\theta _{a},\theta
_{b}$. The Hamiltonian $H(\theta _{a},\theta _{b})$ can be diagonalized by a
standard procedure. And the corresponding ground state is%
\begin{equation}
\left\vert g(\theta _{a},\theta _{b})\right\rangle =R(\theta _{a},\theta
_{b})\left\vert G\right\rangle ,
\end{equation}%
and the bra ground state is
\begin{equation}
\left\langle \overline{g}(\theta _{a},\theta _{b})\right\vert =\left\langle
\overline{G}\right\vert R^{\dag }(\theta _{a},\theta _{b}),
\end{equation}%
where $\theta _{a}$ and $\theta _{b}$ are assumed to be the functions of the
complex field, $\theta _{a,b}=\theta _{a,b}(\eta ,\xi )$. In the following,
we will demonstrate that a appropriate choice of $\theta _{a,b}(\eta ,\xi )$%
\ can connect the geometric phase of the ground state to the boundary of the
phase diagram.

The Berry curvature for the ground state is an anti-symmetric second-rank
tensor derived from the Berry connection via%
\begin{equation}
\mathcal{G}_{\xi \eta }=\frac{\partial }{\partial \xi }\mathcal{B}_{\eta }-%
\frac{\partial }{\partial \eta }\mathcal{B}_{\xi },
\end{equation}%
where%
\begin{eqnarray}
\mathcal{B}_{\lambda } &=&i\sum_{\nu =a,b}\frac{\partial \theta _{\nu }}{%
\partial \lambda }\left\langle \overline{G}\right\vert \sum_{l_{\nu }}\sigma
_{l_{\nu }}^{x}\left\vert G\right\rangle \\
&&+\left\langle \overline{G}\right\vert \left. \partial _{\lambda
}G\right\rangle ,\text{ }\left( \lambda =\eta ,\xi \right) .
\end{eqnarray}%
where we set%
\begin{equation*}
\left\vert \partial _{\lambda }G\right\rangle \equiv \frac{\partial }{%
\partial \lambda }\left\vert G\right\rangle \text{, }\left\langle \partial
_{\lambda }\overline{G}\right\vert \equiv \frac{\partial }{\partial \lambda }%
\left\langle \overline{G}\right\vert
\end{equation*}%
Straightforward derivation shows that%
\begin{equation}
\left\langle \partial _{\xi }\overline{G}\right\vert \left. \partial _{\eta
}G\right\rangle -\left\langle \partial _{\eta }\overline{G}\right\vert
\left. \partial _{\xi }G\right\rangle =0,
\end{equation}%
which indicates that in the case of $\partial \theta _{\nu }/\partial
\lambda =0$, the Berry curvature vanishes. In that case, the adiabatic
evolution along a loop in the $\eta -\xi $ plane is trivial, cannot generate
a nonzero geometric phase. This is what happens for the original Hamiltonian
in Eq. (\ref{H_general}), yielding nothing on the boundary of the phase
diagram from the aspect of geometric phase of the ground state.

As a consequence of the field-dependent phase factor $\theta _{a,b}$, we
have the curvature density $\mathcal{C}=\mathcal{G}_{\xi \eta }/\left(
2N\right) $,
\begin{equation}
\mathcal{C}=\frac{i}{2}\sum_{\nu =a,b}\left( \frac{\partial \theta _{\nu }}{%
\partial \eta }\frac{\partial }{\partial \xi }-\frac{\partial \theta _{\nu }%
}{\partial \xi }\frac{\partial }{\partial \eta }\right) m_{\nu },
\end{equation}%
where%
\begin{equation}
m_{\nu }=\frac{1}{N}\left\langle \overline{G}\right\vert \sum_{l_{\nu
}}\sigma _{l_{\nu }}^{x}\left\vert G\right\rangle ,
\end{equation}%
is defined as the magnetization of sublattice $\nu =a,b$ for the ground
state $\left\vert G\right\rangle $. On the other hand, from the
Hellmann--Feynman theorem, it is easy to obtain%
\begin{eqnarray}
m_{a} &=&\left( m_{b}\right) ^{\ast }=\frac{1}{N}\left\langle \overline{G}%
\right\vert \frac{\partial H}{\partial \left( \eta -i\xi \right) }\left\vert
G\right\rangle \\
&=&\frac{2\partial \varepsilon _{g}}{\partial \eta }+i\frac{2\partial
\varepsilon _{g}}{\partial \xi }.  \notag
\end{eqnarray}%
It is now possible to investigate the physical meaning of the Laplacian of $%
\varepsilon _{g}$. From the definition of $m_{a}$, it is immediate to check
that%
\begin{equation}
\triangledown ^{2}\varepsilon _{g}=\frac{1}{4}\left[ \frac{\partial \left(
m_{a}+m_{b}\right) }{\partial \eta }-i\frac{\partial \left(
m_{a}-m_{b}\right) }{\partial \xi }\right] ,
\end{equation}%
which displays the connection between $\triangledown ^{2}\varepsilon _{g}$\
and the magnetizations.

We now proceed to examine the critical behavior of Berry curvature density $%
\mathcal{G}_{\xi \eta }/\left( 2N\right) $. Unlike $\triangledown
^{2}\varepsilon _{g}$, the result depends on the functions of $\theta _{a,b}$%
. If the phase factors are taken in the simple form%
\begin{equation}
\theta _{a}=\theta _{b}=\eta +\xi ,  \label{function1}
\end{equation}%
the Berry curvature density is explicitly given by

\begin{equation}
\mathcal{C}=i2\left( \frac{\partial ^{2}\varepsilon _{g}}{\partial \xi
\partial \eta }-\frac{\partial ^{2}\varepsilon _{g}}{\partial \eta ^{2}}%
\right) .
\end{equation}%
For this quantity we follow the same steps as in last section. i) The case
of $\eta \neq 0$ and $r^{2}\sim 1$: In the thermodynamic limit, the main
contribution of\ $\mathcal{G}_{\xi \eta }/2N$ near this boundary can be
expressed as%
\begin{equation}
\mathcal{C}\backsim \chi \left( \varphi \right) \int_{0}^{\pi }\digamma
\left( k\right) dk\approx -\frac{\sqrt{2}\chi \left( \varphi \right) }{\pi
\left\vert \cos \varphi \right\vert }\ln \left\vert r-1\right\vert ,
\end{equation}%
where $\chi \left( \varphi \right) =\left[ \sin \left( 2\varphi \right)
-2\cos ^{2}\varphi \right] $. We can see the prefactor $\chi \left( \varphi
\right) /\left\vert \cos \varphi \right\vert $ vanishes at $\varphi =\pi /4$%
, $5\pi /4$, and is discontinuous at $\varphi =\pi /2$, $3\pi /2$. It
indicates that the curvature density $C$ is not divergent at the two
vanishing points.

ii) The case of $\left\vert \xi \right\vert \gg 1$ and $\left\vert \eta
\right\vert \rightarrow 0$. By the similar analysis as above, we have

\begin{equation}
\mathcal{C}\approx -\frac{4\sqrt{2}}{\pi }\int_{0}^{\delta }\frac{1}{%
\epsilon _{1}^{k}}\text{d}k\approx -\frac{4\sqrt{2}}{\pi }\ln \left\vert
\eta \right\vert .
\end{equation}%
We note that the Berry curvature density is divergent at the boundary of the
phase diagram, as what happens in a Hermitian system. However, the Berry
curvature is not an imaginary number as that in a Hermitian system. This is
due to the fact that the evolution\ is non-unitary for a non-Hermitian
system. Nevertheless, the biorthogonal norm is still conserved under the
evolution. It is worth pointing out that the choice of the function in Eq. (%
\ref{function1}) is crucial for the occurrence of the divergence of the
Berry curvature density. For instance, if we take
\begin{equation}
\theta _{a}=-\theta _{b}=\eta +\xi ,  \label{function2}
\end{equation}%
the Berry curvature density has the form $\mathcal{C}=2\left( \partial
^{2}\varepsilon _{g}/\partial \xi \partial \eta -\partial ^{2}\varepsilon
_{g}/\partial \xi ^{2}\right) $, which is divergent on the boundary $r=1$
but not at $\eta =0$.

We perform the numerical simulation of the curvature densities $\mathcal{C}$%
\ for the Hamiltonians with the parameters of Eqs. (\ref{function1}) and (%
\ref{function2}). The shapes of $\mathcal{C}$\ accord with the analytical
predictions in both cases.

\section{Summary}

\label{sec_summary}

In this paper, we explore the QPT in non-Hermitian $\mathcal{PT}$-symmetric
Ising model, which is driven by a staggered complex transverse field. Exact
solution shows that the Laplacian of the groundstate energy density, with
respect to real and imaginary components of the transverse field, diverges
on the boundary in the complex plane. The phase diagram indicates that the
imaginary transverse field has the effect of shrinking the paramagnet phase.
We also investigate the connection between the geometric phase and the QPT\
in the present model as that in the study of the conventional quantum spin
model. We find that the phase boundary can be identified by divergence of
Berry curvature density.

\acknowledgments We acknowledge the support of the National Basic Research
Program (973 Program) of China under Grant No. 2012CB921900 and CNSF (Grant
No. 11374163).

\end{document}